\documentclass[10pt]{article}

\usepackage[T1]{fontenc}
\usepackage[utf8]{inputenc}
\usepackage[english]{babel}
\usepackage{amsmath,latexsym,amsfonts,amsthm,bm,mathtools}
\usepackage{graphicx}
\usepackage[top=1.5in, bottom=1.5in, left=1.25in, right=1.25in]{geometry}
\usepackage{booktabs,url,tikz}
\usepackage[makeroom]{cancel}
\usepackage{dsfont}
\usepackage{relsize}
\usepackage{multirow}

\newcommand{\Varr}[2]{\mathbb{V}\text{ar}_{\tiny{#1}}\left[ #2\right]}

\newcommand{\Expp}[2]{\mathbb{E}_{\tiny{#1}}\left[ #2\right]}

\newcommand{\Prob}[1]{\mathbb{P}(#1)}

\newcommand{\Probb}[2]{\mathbb{P}_{#2}\left(#1\right)}

\newcommand{\approxx}{\overset{\sim}{=}}

\newcommand{\fuzzyset}[1]{\xi_{\widetilde{#1}}}
\newcommand{\indicatorFun}[3]{\mathds{1}_{(#2,#3)}(#1)} 
\newcommand{\digamma}[1]{\Psi\left(#1\right)}

\DeclareMathOperator{\diag}{diag}

\DeclareMathAlphabet{\mathcall}{OMS}{zplm}{m}{n}

\title{{Modeling random and non-random decision uncertainty in ratings data: A fuzzy beta model}}
\author{Antonio Calcagn\`{i}$^{1\ast}$ ~and Luigi Lombardi$^{2}$ \\\\
		\footnotesize{\sl $^{1}$ University of Padova, \sl $^{2}$ University of Trento}\\
		\footnotesize{$\ast$ E-mail: antonio.calcagni@unipd.it}
	}

\date{}

\begin{document}

\maketitle

\begin{abstract}
Modeling human ratings data subject to raters' decision uncertainty is an attractive problem in applied statistics. In view of the complex interplay between emotion and decision making in rating processes, final raters' choices seldom reflect the true underlying raters' responses. Rather, they are imprecisely observed in the sense that they are subject to a non-random component of uncertainty, namely the decision uncertainty. The purpose of this article is to illustrate a statistical approach to analyse ratings data which integrates both random and non-random components of the rating process. In particular, beta fuzzy numbers are used to model raters' non-random decision uncertainty and a variable dispersion beta linear model is instead adopted to model the random counterpart of rating responses. The main idea is to quantify characteristics of latent and non-fuzzy rating responses by means of random observations subject to fuzziness. To do so, a fuzzy version of the Expectation-Maximization algorithm is adopted to both estimate model's parameters and compute their standard errors. Finally, the characteristics of the proposed fuzzy beta model are investigated by means of a simulation study as well as two case studies from behavioral and social contexts. 

\end{abstract}

\vspace{2cm}

\section{Introduction}

In social and behavioral research, satisfaction surveys, aptitude and personality testing, demographic inquiries, and life quality questionnaires are widespread tools to collect data involving subjective evaluations, agreements, and judgments. In a typical social survey, a set of questions is administered to a sample of participants and they are asked to express the extent of their agreement on bounded discrete or continuous rating scales \cite{aiken1996rating,miller2002handbook}. Traditionally, ratings data have been collected by means of pencil and paper questionnaires although more flexible implementations, such as online questionnaires, have become increasingly popular. More recently, technological advancements have fostered the development of new rating tools which offer an accurate way to trace the rating process from its beginning to the final rating outcome \cite{schulte2011handbook}. Hence, unlike standard rating tools, these techniques allow researchers to collect a richer variety of participants' data, including temporal unfolding of ratings and decision uncertainty \cite{calcagni2014dynamic,freeman2010mousetracker}. As several scholars have pointed out, \textit{decision uncertainty} can be referred to a subject-specific cognitive component of the rating process designed to construct a coherent mental representation of the question being rated \cite{ulkumen2016two}. In this sense, it reflects the subjective interplay of decisional and emotional components which contribute to the final rating response \cite{kahneman1982variants}. As such, when appropriately utilized, this source of within-subject heterogeneity can reveal more about ratings then standard crisp responses. Interestingly, this type of non-random, systematic uncertainty has been extensively studied especially with regards to its effects on rating responses (e.g., see \cite{saaleRating1980}). Indeed, it is widely recognized that ratings data often suffer from lack of accuracy, for instance because of social desirability \cite{furnham1986response}, faking behaviors \cite{lombardi2015sgr}, personality \cite{muthukumarana2014bayesian}, response styles \cite{eid2007detecting}, and violations of rating rules \cite{iannario2015modelling,preston2000optimal,rabinowitz2019consistency}. These issues have not only been recognized as important by applied statisticians working with ratings data but also by several researchers working in fields like applied econometrics (e.g., see \cite{angel2019did,de2011measuring,zafar2011can}), metrology (e.g., see \cite{pendrill2016metrology,pendrill2014using}), and risk analysis (e.g., see \cite{slovic2004risk}). 

Modeling ratings data is a relevant problem in applied statistics. Commonly used methods to handle with discrete or continuous rating data include generalized linear models (GLMs) \cite{mcculloch2000generalized}, beta regression models \cite{ferrari2004beta,migliorati2018new,ospina2012general}, and combination of uniform and shifted binomial (CUBE) models  \cite{golia2015interpretation,piccolo2019class,piccolo2019cumulative}. These models typically represent mean and dispersion components as a linear or non-linear function of some external covariates, which are intended to explain the observed heterogeneity of the ratings data. Although some of these methods also allow for disentangling individual indecision and heterogeneity of responses induced by the presence of subgroups (e.g., CUBE), they are mainly intended to work with data represented as a crisp collection of responses and do not account for non-random decision components of rating process. The same applies with more general approaches to analyse within-subject heterogeneity such as random-effects and errors-in-variables models \cite{feng2018statistical} which do not deal with non-random components of uncertainty. As a consequence, decision uncertainty underlying participant's rating process is not formally represented in these models.

In this contribution we propose a novel method for analysing continuous bounded ratings data that are characterized by non-random and systematic decision uncertainty. In particular, we propose a variable dispersion beta linear model which is generalized to cope with data contaminated by subjective uncertainty. We represent decision uncertainty in the framework of fuzzy data modeling, where crisp ratings data are equipped with non-random systematic uncertainty via normalized set functions \cite{couso2014statistical,kruse1987statistics}. In this setting, maximum likelihood estimation and inference are carried out through the Expectation-Maximization algorithm adapted for the case of fuzzy data \cite{denoeux2011maximum,su2015parameter}. It should be stressed that using beta linear models allows for flexibility in modeling and analysing continuous ratings data, while still retaining simplicity in estimates model's parameters \cite{algamal2019particle,canterle2019variable,zeileis2010beta}. Similarly, despite the fact that many formal theories have been proposed to deal with subjective uncertainty (e.g., soft sets, rough sets. See \cite{lin2012rough,liu2010uncertainty}), fuzzy set theory offers a good compromise in terms of accuracy and computational costs and benefit from a long tradition of works in statistics (e.g., see \cite{buckley2006fuzzy,chukhrova2018randomized,couso2014statistical,gebhardt1998fuzzy,gonzalez2006fuzzy}).

The remainder of the article is organized as follows. Section 2 briefly describes the basic characteristics of fuzzy data together with their interpretation in terms of decision uncertainty. Section 3 exposes the variable dispersion fuzzy beta model, parameters estimation, and model evaluation. Section 4 reports results of a short simulation study performed to evaluate the finite sample properties of the fuzzy beta model and the consequences of neglecting decision uncertainty on parameters estimation. Section 5 describes an application of the new approach to two case studies involving ratings data from risk-taking behaviors (application 1) and customer satisfaction (application 2). Finally, Section 6 concludes the article providing final remarks and suggestions for future extensions.  All the materials like datasets and \texttt{R}-scripts used throughout the article are available to download at \url{https://github.com/antcalcagni/fuzzyratingbeta}.

\subsection{Fuzzy numbers and fuzzy probability}

A fuzzy set $\tilde{A}$ of a universal set $\mathcall A$ is defined by means of its characteristic function $\fuzzyset{A}:\mathcall{A}\to [0,1]$. It can be easily described as a collection of crisp subsets called $\alpha$-sets, i.e. $\tilde{A}_\alpha = \{x \in \mathcall A: ~\fuzzyset{A}(x) > \alpha \}$ with $\alpha \in (0,1]$. If the $\alpha$-sets of $\tilde{A}$ are all convex sets then $\tilde{A}$ is a {convex fuzzy set}. The {support} of $\tilde{A}$ is $\tilde{A}_{0} = \{x \in \mathcall A: ~\fuzzyset{A}(x) > 0 \}$ and the core is the set of all its maximal points, i.e. $\tilde{A}_{c} = \{x \in \mathcall A: ~\fuzzyset{A}(x) = \max_{y \in \mathcall A}~ \fuzzyset{A}(y) \}$. In the case $\max_{x\in \mathcall A} \fuzzyset{A}(x) = 1$ then $\tilde{A}$ is a {normal} fuzzy set. If $\tilde{A}$ is a normal and convex subset of $\mathbb R$ then $\tilde{A}$ is a fuzzy number \cite{buckley2006fuzzy}. Broadly speaking, fuzzy sets can be conceived as subsets of $\mathbb{R}$ where their Boolean characteristic function $\xi_{A}(x)$, $\forall x\in A \subset \mathbb{R}$, has been generalized to the real interval $[0,1]$. The class of all normal fuzzy numbers is denoted by $\mathcall F(\mathbb{R})$. Fuzzy numbers can conveniently be represented using parametric models, through which $\fuzzyset{A}$ is represented by means of few real parameters. Hence, we can define families of fuzzy numbers indexed by some scalars, such as $m$ (mode) and $s$ (spread/precision), which include a number of shapes like triangular, trapezoidal, gaussian, and exponential \cite{buckley2006fuzzy}. Relevant classes of parametric fuzzy numbers are the so-called LR-fuzzy numbers \cite{dubois1978operations} and their generalizations \cite{calcagni2014non,dombi2018flexible}. In this setting, a set of operators and algebras have also been defined for fuzzy numbers, which extend traditional calculus to fuzzy numbers as well \cite{chwastyk2013fuzzy}. A broader class that encompasses a wide range of fuzzy numbers is the so-called \textit{beta fuzzy number} \cite{alimi2003beta,baklouti2018beta,stein1985fuzzy}:
\begin{align}\label{eq1}
	& \fuzzyset{A}(x) = \bigg(\frac{x-x_l}{m-x_l}\bigg)^a \bigg(\frac{x_u-x}{x_u-m}\bigg)^b \cdot \indicatorFun{x}{x_l}{x_u}\\
	& m = \frac{ax_u + bx_l}{a+b} \nonumber
\end{align} 
where $x_l, x_u, a,b \in \mathbb R$, with $x_l$ and $x_u$ being the lower and upper bounds of the set, and $m$ the mode of the fuzzy set. This type of fuzzy numbers uses Beta functions to approximate many regular shapes such as triangular, trapezoidal or Gaussian. Likewise for LR-fuzzy numbers, beta fuzzy numbers can be defined in terms of mode $m \in \mathbb R$ and spread/precision $s \in \mathbb{R}^+$ parameters. In particular, let $x_l=0$ and $x_u=1$ without loss of generality. Then Eq. \eqref{eq1} can be re-arranged as follows:
\begin{align}\label{eq2}
	&\fuzzyset{A}(x) = \frac{1}{C}~ x^{a-1} (1-x)^{b-1}\\
	& a = 1+ms\nonumber\\
	& b = 1+s(1-m)\nonumber
\end{align} 
with $C$ being a constant ensuring $\fuzzyset{A}$ is still a normal fuzzy set:
\begin{equation*}
	C = \bigg(\frac{a-1}{a+b-2}\bigg)^{a-1} \cdot~ \bigg(1-\frac{a-1}{a+b-2}\bigg)^{b-1}
\end{equation*}
Interestingly, the re-parameterized Beta fuzzy number resembles the shape of Beta density distribution written using the PERT representation \cite{vose2008risk}. {It should be noted that, like triangular or trapezoidal fuzzy numbers, also beta fuzzy numbers can be adopted to represent rating data variables. Indeed, the beta-like shape is particularly flexible to modeling some features of bounded rating data, such as asymmetry and tail flatness (e.g., see \cite{migliorati2018new}).}

When a probability space is defined over the reals, the probability of a fuzzy set $\Prob{\tilde{A}}$ can also be defined. Over the years, there have been various attempts to define the probability of a fuzzy set in terms of expected value of its membership function \cite{zadeh1968probability}, conditional probability of prior information \cite{coletti2004conditional,singpurwalla2004membership}, imprecise probability \cite{dubois2010imprecise}, fuzzy numbers \cite{hesamian2017note}, and likelihood induced by random events \cite{cattaneo2017likelihood}. Following the findings of \cite{denoeux2011maximum}, in this contribution we adopt Zadeh's definition of fuzzy probability \cite{zadeh1968probability}. In particular, let $(\mathbb R,\mathcall B, \mathbb P)$ be a probability space. Then, $\Prob{\tilde{A}}$ is defined as follows:
\begin{equation}\label{eq3a}
	\Prob{\tilde A} = \int_{\mathbb{R}} \xi_{\tilde A}(x) d\mathbb P	
\end{equation}
with $\fuzzyset{A}$ being Borel measurable. In this context, two fuzzy sets $\tilde{A}$ and $\tilde{B}$ are said \textit{independent} w.r.t. to $\mathbb P$, if $\Prob{\tilde A \tilde B} = \Prob{\tilde{A}}\cdot\Prob{\tilde B}$, with the fuzzy product being defined as $\xi_{\tilde A \tilde B}(x) = \xi_{\tilde A}(x)\cdot \xi_{\tilde B}(x)$ \cite{zadeh1968probability}. The \textit{conditional probability} of two independent fuzzy events is 
\begin{equation}\label{eq3b}
	\Prob{\tilde A|\tilde{B}} = \frac{\Prob{\tilde{A},\tilde{B}}}{\Prob{\tilde{B}}} = \frac{\int \xi_{\tilde{A}}(x)\xi_{\tilde B}(x) d~\mathbb P(x)}{\int \xi_{\tilde B}(z) d~\mathbb P(z)}
\end{equation}
with $\Prob{\tilde{B}} > 0$. Note that one can also obtain the conditional probability between a crisp set $A$ and a fuzzy set $\tilde{B}$ as a special case of Eq. \eqref{eq3b}. If a discrete or continuous random variable $Y$ is defined over $\mathcall B$, then fuzzy probability can be generalized accordingly. For instance, denoting with $f_Y(y;\boldsymbol{\theta})$ the probability density of $Y$, then $\Prob{\tilde A} = \int_{\mathcall {Y}} \fuzzyset{A}(y) f_Y(y;\boldsymbol{\theta}) ~dy$, with $\mathcall Y$ being the support of $Y$. Similarly, when a sample of $n$ independent observations $\mathbf y = (y_1,	\ldots,y_n)$ from $Y_1,\ldots,Y_n$ is available, the likelihood of the sample $\mathbf{y}$ can be generalized as follows:
\begin{equation}\label{eq4}
	\mathcall L(\mathbf y;\boldsymbol{\theta}) = \prod_{i=1}^{n} \int_{\mathcall {Y}} \fuzzyset{y_i}(y) ~f_{Y_i}(y;\boldsymbol{\theta}) ~dy
\end{equation}
where the definition $\fuzzyset{\mathbf y} = \prod_{i=1}^n \fuzzyset{y_i}(y)$ has been used for the joint fuzzy set \cite{gebhardt1998fuzzy}. Further details about fuzzy generalization of likelihood functions, fuzzy random variables, and fuzzy probability space can be found in \cite{cattaneo2017likelihood,couso2014random,denoeux2011maximum,gebhardt1998fuzzy,gil2006overview}.

We interpret fuzzy data in the context of random variables following the epistemic viewpoint on fuzzy set theory \cite{couso2014statistical}. In particular, for a fuzzy set $\tilde{A}$, $\fuzzyset{A}(Y=y)$ is interpreted as the \textit{possibility} that the crisp event $Y=y$ has to occur. Indeed, $\fuzzyset{A}(Y=y) \in (0,1)$ can be conceived as a graded plausibility about the occurrence of the event $Y=y$, with $\fuzzyset{A}(Y=y) = 1$ indicating the fact that $Y=y$ is fully possible. By contrast, $\fuzzyset{A}(Y=y) = 0$ indicates that $Y=y$ is not possible at all. Hence, fuzzy sets can intuitively be viewed as \textit{graded constraints} on crisp random variables. In this way, the {randomness} due to the data generation process and the {fuzziness} due to observer's state of knowledge can be analysed simultaneously by means of a common statistical representation. As a remark, note that in this setting $\fuzzyset{A}(Y=y)$ is thought as being the consequence of a two-step generation process, in which a realization $y$ is drawn from $Y$ first and then a fuzzy set $\fuzzyset{A}$ is used to encapsulate the uncertainty about $Y=y$ in terms of possibility distribution. Hence, only the first stage is a random experiment whereas the second stage is a non-random fuzzification of the outcomes being realized.

\subsection{Fuzzy scaling and fuzzy data} 

Since the seminal work of \cite{hesketh1988application}, fuzzy sets have been extensively used in the context of ratings data (for a review, see \cite{calcagni2014dynamic,lubiano2016descriptive,calcagn2021psychometric}). Although several formats have been proposed for implementing fuzzy ratings tools (e.g., conversion scales, direct rating scales, implicit rating scales), all of them share the same idea that ratings data cannot be coerced into crisp numbers without a certain loss of information. Except for the case of dichotomous ratings, polytomous responses often show some degrees of imprecision and fuzziness, which is essentially due to participants' rating processes \cite{kahneman1982variants}. In general, ratings data can be enriched by including information related to participants' response process and {fuzzy conversion} or {fuzzy rating} systems can be adopted to this purpose. {In what follows, we will briefly review both the approaches to fuzzy rating.}

{\textit{Fuzzy conversion scales} aim at turning standard crisp ratings into fuzzy numbers through the adoption of user-defined fuzzy systems (e.g., see \cite{vonglao2017application}) or statistically-oriented procedures (e.g., see \cite{yu2009fuzzy}). In general, a typical implementation of a user-defined fuzzy conversion scale, a fuzzy system relates a space of crisp responses (input) to a space of fuzzy numbers (output). Next implication rules mapping both input and output sources are finally established. For instance, in the simplest case of a 5-point scale, the crisp response Y = 2 activates the fuzzy sets of the output space via an IF-THEN implication rule. On the basis of the implemented fuzzy system  (e.g., Mamdani, Sugeno), the output can result in the activation of one or two fuzzy sets and the final fuzzy response - which is given by the intersection of the activated fuzzy sets - will represent a more certain or less certain response, respectively.}

{\textit{Fuzzy rating scales} adopt computerized interfaces through which the rater's response process is directly mapped to fuzzy numbers (e.g., see \cite{hesketh1988application,calcagni2014dynamic,de2014fuzzy}). For instance, in the direct rating based implementation \cite{de2014fuzzy}, raters are asked to respond by drawing fuzzy sets according to their perceived uncertainty. In this case, the rating procedure proceeds as follows. First, raters draw an interval on a pseudo-continuous graphical scale, which represents the set of admissible responses compatible with the assessment of the item being rated. Then, a degree of confidence is expressed by drawing another interval around the previous interval response. Finally, both the intervals are interpolated to form the final fuzzy responses (e.g., trapezoidal or triangular). In a similar way, fuzzy indirect scales adopt a computerized graphical interface to get responses from raters. However, in this case, they are not asked to directly draw their responses; rather, fuzzy sets are build from a set of implicit measures associated to the final crisp response (e.g., response time). For instance, the DYFRAT scale \cite{calcagni2014dynamic} uses a set of biometric measures associated with the cognitive response process (i.e., response time, computer-mouse trajectories) to derive the rater's fuzzy response. In particular, raters are presented with a pseudo-circular scale with $M$ levels and - similarly to the case of crisp Likert-type scales - they are asked to choose which of these levels is the most appropriate to represent their response for a given item being rated. Meanwhile, the system records the streaming coordinates of the computer mouse cursor (at a fixed sampling rate) and the elapsed time since the beginning of the response trajectory. Finally, both information are used to define a fuzzy response according to the following rationale: (i) computer-mouse trajectories are projected onto a linear scale and the histogram of linearized radians is used to derive the fuzzy set (e.g., triangular, beta) in terms of its support and core, (ii) response times are used to intensify or reduce the fuzziness of the sets by means of linguistic quantifiers (e.g., the longer the response time, the higher the fuzziness of the set).}

{In the context of fuzzy scaling, fuzzy numbers are used as a formal representation for rating responses involving individual-based judgments, attitudes, and opinions. Although fuzzy conversion and rating scales differ in the way they derive fuzzy responses, both aim at providing a model for the fuzziness or imprecision which is present in the rating process $Y\sim f_Y(y;\boldsymbol{\theta})$. As for the more general case of LR fuzzy numbers, the parameters of a beta fuzzy number can be linked to the rating process as follows.} First, consider a continuous rating scale bounded on a subset $(y_l,y_u)$ of reals. Then, $m$ represents the most plausible final rating choice $\fuzzyset{Y}(Y=y) = 1$, $s$ is the precision of $m$ such that smaller values indicate larger levels of hesitation in the rating choice, and $\fuzzyset{Y}$ conveys the overall decision uncertainty in terms of fuzziness (the larger the fuzziness, the highest the decision uncertainty). Note that, ideally, if there was no subjective uncertainty, then the fuzziness would tend to zero and true rating realizations would be precisely observed (i.e., $Y=y=m$). In this case, there would not need to represent ratings as fuzzy data.

\section{Variable dispersion beta model for fuzzy ratings}

In this section we illustrate our proposal to analyse continuous bounded ratings data in situations with decision uncertainty. Hereafter, ratings data will be considered scaled into the real subset $(0,1)$ without loss of generality.

\subsection{Model}

Let $\mathbf y = (y_1,\ldots,y_i,\ldots,y_n) \in (0,1)^n$ be a sample of $n$ observations from Beta distributed independent random variables $Y_1,\ldots,Y_i,\ldots,Y_n$ with density:
\begin{equation}\label{eq5}
	f_{\mathbf Y}(\mathbf y;\boldsymbol\mu,\boldsymbol\phi) = \prod_{i=1}^n ~ \frac{\Gamma(\phi_i)}{\Gamma(\phi_i\mu_i)\Gamma(\phi_i-\mu_i\phi_i)}~y_i^{(\mu_i\phi_i-1)} (1-y_i)^{(\phi_i-\mu_i\phi_i-1)}
\end{equation}
with $\boldsymbol\mu \in (0,1)^n$ being the $n\times 1$ vector of location parameters and $\boldsymbol\phi \in (0,\infty)^n$ the $n\times 1$ vector of precision parameters \cite{ferrari2004beta}. The sequence $Y_1,\ldots,Y_i,\ldots,Y_n$ models the ratings for each of the $n$ participants, with the convention that $\{0,1\}$ represent the lower and upper bounds of the rating domain, respectively. In order to account for heterogeneity and non-constant variance in rating responses, location and dispersion parameters can be non-linearly re-written using monotonic and twice differentiable link functions, mapping the support $(0,1)$ into $\mathbb R$, as follows:
\begin{equation}\label{eq6}
	g_1(\boldsymbol\mu) = \mathbf X\boldsymbol\beta \quad\text{and}\quad g_2(\boldsymbol\phi) = \mathbf Z\boldsymbol\gamma
\end{equation}
where $\mathbf X$ and $\mathbf Z$ are $n\times J$ and $n\times H$ matrices of known continuous or categorical covariates, with $\boldsymbol{\beta}$ and $\boldsymbol{\gamma}$ being vectors of appropriate order containing unknown parameters. The functions $g_1(.)$ and $g_2(.)$ can be chosen among a variety of link functions (e.g., logit, probit, log. See \cite{mcculloch2000generalized}). Two typical choices are the logit and logarithm functions, which yield to:
\begin{equation}\label{eq6b}
	\boldsymbol\mu = \big({1+\exp(\mathbf X\boldsymbol\beta)}\big)^{-1} \quad\text{and}\quad \boldsymbol\phi = \exp(\mathbf Z\boldsymbol\gamma)
\end{equation}
Under Eq. \eqref{eq6b}, the log-likelihood function for the variable dispersion beta model is:
\begin{equation}\label{eq7}
	\begin{split}
		l(\boldsymbol{\theta};\mathbf y) =& \sum_{i=1}^n ~\ln\Gamma\big(\exp(\mathbf{z}_i\boldsymbol{\gamma})\big) - \ln\Gamma\left(\frac{\exp(\mathbf{z}_i\boldsymbol{\gamma})}{1+\exp(-\mathbf x_i\boldsymbol{\beta})}\right) - \\  & \ln\Gamma\left(\exp(\mathbf{z}_i\boldsymbol{\gamma})-\frac{\exp(\mathbf{z}_i\boldsymbol{\gamma})}{1+\exp(-\mathbf x_i\boldsymbol{\beta})}\right) +\\
		& \ln y_i\left( \frac{\exp(\mathbf z_i\boldsymbol{\gamma})}{1+\exp(-\mathbf x_i\boldsymbol{\beta} )} - 1\right) + \\ 
		&\ln (1-y_i)\left(\exp(\mathbf{z}_i\boldsymbol{\gamma})-\frac{\exp(\mathbf{z}_i\boldsymbol{\gamma})}{1+\exp(-\mathbf x_i\boldsymbol{\beta})} - 1\right)
	\end{split}
\end{equation}
with $\Gamma(.)$ being the Euler gamma function and $\mathbf t = \big(\ln \mathbf y, \ln \mathbf (\mathbf 1 - \mathbf y)\big)$ the sufficient statistics for the inference on $\boldsymbol{\theta} = (\boldsymbol{\beta},\boldsymbol{\gamma})$. 

In light of the data representation adopted in this work, decision uncertainty is treated as a systematic and non-random component which occurs after the sampling process $\mathbf y\sim f_{\mathbf Y}(\mathbf y;\boldsymbol\mu,\boldsymbol\phi)$ has been realized. This leads to a situation where the sample $\mathbf y$ cannot be precisely observed and a collection of fuzzy data $\mathbf{\tilde{y}}$ is instead available. When fuzzy data are represented as beta fuzzy numbers, then 
\begin{equation*}
	\mathbf{\tilde{y}} = (\mathbf m,\mathbf s) = \big((m_1,s_1),\ldots,(m_i,s_i),\ldots,(m_n,s_n)\big)	
\end{equation*}
with $\mathbf m$ and $\mathbf s$ being $n\times 1$ vectors of modes and precisions/spreads for the fuzzy observations. Turning Eq. \eqref{eq3a} into \eqref{eq5}, the joint density of $\mathbf{\tilde y}$ can be written as:
\begin{align}\label{eq8}
	f_{\mathbf{\widetilde Y}}(\mathbf{y};\boldsymbol\mu,\boldsymbol\phi) &= \prod_{i=1}^n \int_{\mathcall {Y}} \fuzzyset{{y}_i}(y) ~d\Probb{Y_i=y}{\boldsymbol{\theta}}\nonumber\\
	&= \prod_{i=1}^n \int_{\mathcall {Y}} \fuzzyset{{y}_i}(y)~\frac{\Gamma(\phi_i)y^{(\mu_i\phi_i-1)} (1-y)^{(\phi_i-\mu_i\phi_i-1)}}{\Gamma(\phi_i\mu_i)\Gamma(\phi_i-\mu_i\phi_i)} ~dy
\end{align}
where the joint fuzzy set $\fuzzyset{\mathbf y} = \prod_{i=1}^n \fuzzyset{y_i}(y)$ has been factorized in terms of product. The log-likelihood of the model under fuzzy observations can analogously be obtained using Eq. \eqref{eq8}. Note that, in this representation the vectors $\mathbf m$ and $\mathbf s$ of the fuzzy sets enter the model as observed quantities whereas the parameters $\boldsymbol{\beta}$ and $\boldsymbol{\gamma}$ still remain non-fuzzy quantities.

\subsection{Parameter estimation} 

To provide estimates for $\boldsymbol{\theta} = (\boldsymbol{\beta},\boldsymbol{\gamma})$ in the context of fuzzy ratings data, one can maximize the log-likelihood $l(\boldsymbol{\theta};\mathbf{\tilde y)}$ function, which is obtained by Eq. \eqref{eq8}. This would require an iterative procedure, alternating between the numerical computation of the integral and the maximization of the function. However, to avoid the problem of approximating integrals in Eq. \eqref{eq8} and have a way to compute standard errors consistently, we will use a variant of the Expectation-Maximization algorithm generalized for the case of fuzzy data \cite{denoeux2011maximum}. As for the standard EM algorithm, the fuzzy-EM version at the $k$-th iteration alternates between the E-step, which involves the computation of the expected complete-data log-likelihood using $\boldsymbol{\theta}^{(k-1)}$, and the M-step, which instead maximizes the expected complete-data log-likelihood w.r.t. to $\boldsymbol{\theta}^{(k)}$. These steps generate a non-decreasing sequence of lower bounds for the maximization of the observed-data log-likelihood $l(\boldsymbol{\theta};\mathbf{\tilde y)}$ (for formal details, see \cite{denoeux2011maximum}). In the fuzzy-EM variant, the complete-data log-likelihood is that obtained if $\mathbf y$ was precisely observed (see Eq. \ref{eq7}). Therefore, given the $(k-1)$-th estimates $\boldsymbol{\theta}' = \boldsymbol{\theta}^{(k-1)}$ the E-step for the $k$-th iteration of the algorithm consists in the computation of the following quantity:
\begin{align}\label{eq9}
	\mathcall Q\left(\boldsymbol\theta,\boldsymbol\theta'\right) & = \Expp{\boldsymbol{\theta}'}{l(\boldsymbol{\theta};\mathbf y)\big|\mathbf{\tilde y}}\nonumber\\
	& = C + \sum_{i=1}^n ~\Expp{\boldsymbol{\theta}'}{\ln Y_i\big|{\tilde y_i}}\left( \frac{\exp\left(\mathbf z_i\boldsymbol{\gamma}'\right)}{1+\exp\left(-\mathbf x_i\boldsymbol{\beta}' \right)} - 1\right) + \nonumber\\
	&  \sum_{i=1}^n ~\Expp{\boldsymbol{\theta}'}{\ln (1-Y_i)\big|{\tilde y_i}}\left(\exp\left(\mathbf{z}_i\boldsymbol{\gamma}'\right)-\frac{\exp\left(\mathbf{z}_i\boldsymbol{\gamma}'\right)}{1+\exp\left(-\mathbf x_i\boldsymbol{\beta}'\right)} - 1\right)
\end{align}
where $C$ contains all the terms of Eq. \eqref{eq7} that do not involve the random quantities to be filtered. To compute the conditional expectations of the E-step, note that $Y_i\big|{\tilde y_i}$ is a random variable conditioned on fuzzy events and its density can be obtained using Eq. \eqref{eq3b} simplified for the case where $A$ is a crisp event:
\begin{align}\label{eq10}
	f_{Y_i|\tilde{y}_i}(y;\mu_i,\phi_i) &= \frac{\fuzzyset{y_i}(y) f_{Y_i}(y;\mu_i,\phi_i)}{\int_{\mathcall Y} \fuzzyset{y_i}(x) f_{Y_i}(x;\mu_i,\phi_i) ~dx}\nonumber\\
	& \propto y_i^{(\eta_i - 1)} (1-y_i)^{(\nu_i -1)}
\end{align}
Under the case of beta fuzzy numbers and up to a normalization constant, the conditional density $f_{Y_i|\tilde{y}_i}(y;\mu_i,\phi_i)$ corresponds to a beta density with parameters given as a function of fuzzy data and crisp parameters of the complete-data density:
\begin{align}\label{eq10b}
	& \eta_i = \mu_i\phi_i + s_im_i + 1\nonumber\\
	& \nu_i = \phi_i(1-\mu_i) + s_i(1-m_i) +1
\end{align}  
Then, the first expectation $\Expp{\boldsymbol{\theta}'}{\ln Y_i\big|{\tilde y_i}}$ can be approximated via Taylor expansion around $\Expp{\boldsymbol{\theta}'}{Y_i\big|{\tilde y_i}}$ as follows:
\begin{align}\label{eq11a}
	\Expp{\boldsymbol{\theta}'}{\ln Y_i\big|{\tilde y_i}} ~\approxx~& \ln ~\Expp{\boldsymbol{\theta}'}{Y_i\big|{\tilde y_i}} - \frac{1}{2}\frac{\Varr{\boldsymbol{\theta}'}{Y_i|\tilde{y}_i}}{\left(\Expp{\boldsymbol{\theta}'}{Y_i|\tilde{y}_i}\right)^2}\nonumber\\
	~\approxx~& \ln \left(\frac{\eta_i}{\eta_i+\nu_i}\right) - \frac{\nu_i}{2\eta_i(1+\eta_i+\nu_i)}
\end{align}
Similarly, the second expectation $\Expp{\boldsymbol{\theta}'}{\ln (1-Y_i)\big|{\tilde y_i}}$ is obtained by symmetry of the beta function:
\begin{align}\label{eq11b}
	\Expp{\boldsymbol{\theta}'}{\ln (1-Y_i)\big|{\tilde y_i}} ~\approxx~& \ln ~\Expp{\boldsymbol{\theta}'}{(1-Y_i)\big|{\tilde y_i}} - \frac{1}{2}\frac{\Varr{\boldsymbol{\theta}'}{(1-Y_i)|\tilde{y}_i}}{\left(\Expp{\boldsymbol{\theta}'}{(1-Y_i)|\tilde{y}_i}\right)^2}\nonumber\\
	~\approxx~& \ln \left(\frac{\nu_i}{\eta_i+\nu_i}\right) - \frac{\eta_i}{2\nu_i(1+\eta_i+\nu_i)}
\end{align}

Once expected values are computed, the M-step of the algorithm involves the maximization of $\mathcall Q\left(\boldsymbol\theta,\boldsymbol\theta'\right)$ with respect to the elements of $\boldsymbol{\theta}$ and can be performed by plugging-in Eqs. \eqref{eq11a}-\eqref{eq11b} into the log-likelihood Eq. \eqref{eq9}. The simultaneous score equations for M-step are as follows:
\begin{align}\label{eq12a}
	\mathcall U_{\boldsymbol{\beta}} &= \frac{\partial \mathcall Q\left(\boldsymbol\theta,\boldsymbol\theta'\right)}{\partial \boldsymbol{\mu}}\frac{\partial \boldsymbol{\mu}}{\partial\boldsymbol{\beta}} = \mathbf 0_J\nonumber\\
	&= \mathbf X^T \big[ \boldsymbol{\phi}\big(\mathbf y^*_{(1)} - \mathbf y^*_{(2)} - 
	\digamma{\boldsymbol{\phi}\boldsymbol{\mu}} + \digamma{\boldsymbol{\phi}(\mathbf 1-\boldsymbol{\mu})} \big)\big]\boldsymbol{\mu}^2 = \mathbf 0_J
\end{align}
\begin{align}\label{eq12b}
	\mathcall U_{\boldsymbol{\gamma}} &= \frac{\partial \mathcall Q\left(\boldsymbol\theta,\boldsymbol\theta'\right)}{\partial \boldsymbol{\phi}}\frac{\partial \boldsymbol{\phi}}{\partial\boldsymbol{\gamma}} = \mathbf 0_H\nonumber\\
	&= \mathbf Z^T \big[ \boldsymbol{\phi}\digamma{\boldsymbol{\phi}} - \digamma{\boldsymbol{\phi} - \boldsymbol{\phi}\boldsymbol{\mu}}\boldsymbol{\phi} + \digamma{\boldsymbol{\phi} - \boldsymbol{\phi}\boldsymbol{\mu}}\boldsymbol{\phi}\boldsymbol{\mu} + \mathbf y^*_{(1)}\boldsymbol{\mu}\boldsymbol{\phi} - \nonumber\\
	& - \boldsymbol{\phi}\digamma{\boldsymbol{\mu}\boldsymbol{\phi}}\boldsymbol{\mu} - \mathbf y^*_{(2)}\boldsymbol{\phi}(\boldsymbol{\mu}-\mathbf 1) \big] = \mathbf 0_H
\end{align}
where $\boldsymbol{\mu}$ and $\boldsymbol{\phi}$ are defined as in Eq. \eqref{eq6b} whereas the terms
\begin{align*}
	& \mathbf y^*_{(1)} = \left(\Expp{\boldsymbol{\theta}'}{\ln Y_1\big|{\tilde y_1}},\ldots,\Expp{\boldsymbol{\theta}'}{\ln Y_n\big|{\tilde y_n}}\right)\\
	& \mathbf y^*_{(2)} = \left(\Expp{\boldsymbol{\theta}'}{\ln (1-Y_1)\big|{\tilde y_1}},\ldots,\Expp{\boldsymbol{\theta}'}{\ln (1-Y_n)\big|{\tilde y_n}}\right)
\end{align*}
denote the filtered data which are computed using Eqs. \eqref{eq11a}-\eqref{eq11b}. Finally, $\boldsymbol{\theta}^{(k)}$ is obtained by computing the roots of Eqs. \eqref{eq12a}-\eqref{eq12b} numerically, for instance using Gauss-Newton method.

\subsection{Standard errors, diagnostics, and inference} 

Standard errors for $\boldsymbol{\hat\beta}$ and $\boldsymbol{\hat\gamma}$ can be computed using the observed information matrix $\mathcall I_{\boldsymbol{\hat\theta}}$ from the Hessian of the maximum likelihood estimates $\boldsymbol{\hat\theta}$ obtained solving Eqs. \eqref{eq12a}-\eqref{eq12b} for $\boldsymbol{\beta}$ and $\boldsymbol{\gamma}$. In the context of EM algorithm, $\mathcall I_{\boldsymbol{\hat\theta}}$ can be approximated using the empirical observed information matrix \cite{meilijson1989fast}, as follows:
\begin{equation}\label{eq13}
	\mathcall I_{\boldsymbol{\hat\theta}} ~\approxx~ \mathcall I^e_{\boldsymbol{\hat\theta}} = \sum_{i=1}^n ~\mathcall U_{\boldsymbol{\hat\theta}}^{(i)} \left(\mathcall U_{\boldsymbol{\hat\theta}}^{(i)}\right)^T
\end{equation} 
where $\mathcall U_{\boldsymbol{\hat\theta}}^{(i)} = [\mathcall U_{\boldsymbol{\hat\beta}}^{(i)};\mathcall U_{\boldsymbol{\hat\gamma}}^{(i)}]$ is the score vector for the $i$-th observation calculated at $\boldsymbol{\hat\beta}$ and $\boldsymbol{\hat\gamma}$. The standard errors are then calculated as usual:
\begin{equation}\label{eq14}
	\sigma_{\boldsymbol{\hat\theta}} = (\diag\{(\mathcall I_{\boldsymbol{\hat\theta}}^e)^{-1}\})^{\frac{1}{2}}
\end{equation} 
Note that this approximation avoids the computation of Hessian of the complete-data log-likelihood as it solely uses the score equations where the unobserved vector $\mathbf y$ is replaced by $\mathbf y^*_{(1)}$ and $\mathbf y^*_{(2)}$. Alternatively, standard errors can also be obtained via non-parametric bootstrap \cite{mclachlan2004finite}. An important quantity commonly used to assess the quality of the estimated model is that involving standardized residuals which, for the case of fuzzy data, can be generalized as follows:
\begin{equation}\label{eq15}
	r_i = \frac{(m_i-\hat\mu_i)(1-\fuzzyset{y_i}(\hat\mu_i))}{{\hat\mu}_i(1 - {\hat\mu}_i) / (1+{\hat\phi}_i)}
\end{equation}
where $\hat{\mu}_i = (1+\exp(\mathbf x_i\boldsymbol{\hat\beta}))^{-1}$ and $\hat{\phi} = \exp(\mathbf z_i\boldsymbol{\hat\gamma})$ whereas $\fuzzyset{y_i}(\hat\mu_i)$ is the fuzzy membership computed for the predicted quantity $\hat\mu_i$. In general, diagnostics for the model can be performed by plotting, for instance, $r_1,\ldots,r_n$ against the indices of the observations in order to check for particular trends or patterns in the predicted data. {Similarly to generalized and beta linear models, also the overall fit of the estimated fuzzy beta model with respect to the observed fuzzy data can be assessed by means of pseudo-$R^2$ indices, which generalize the standard residuals-based $R^2$ indices \cite{ferrari2004beta,veall1994evaluating}. In this context, we resorted in applying a likelihood-ratio based pseudo-$R^2$ index \cite{aldrich1984linear,veall1994evaluating} which, like for the McFadden's pseudo-$R^2$ \cite{mcfadden1973conditional}, contrasts the likelihood value $l(\boldsymbol{\hat\theta};\mathbf y)_{\mathcall{M}_1}$ of the fuzzy beta linear model against the likelihood value of the null model $l(\boldsymbol{\hat\theta};\mathbf y)_{\mathcall{M}_0}$ (i.e., a fuzzy beta linear model with no predictors for $\boldsymbol{\mu}$ and $\boldsymbol{\phi}$). The index is as follows:
	\begin{equation*}
		\text{pseudo-}R^2 =  -\frac{\omega(1-\lambda)}{(\omega+n)\lambda}
	\end{equation*}
	where $\omega = 2\left(l(\boldsymbol{\hat\theta};\mathbf y)_{\mathcall{M}_1}-l(\boldsymbol{\hat\theta};\mathbf y)_{\mathcall{M}_0}\right)$ and $\lambda = \frac{1}{n}l(\boldsymbol{\hat\theta};\mathbf y)_{\mathcall{M}_0}$. The likelihood-ratio based pseudo-$R^2$ is normalized in $[0,1]$ and approximates the relationship between the likelihood-ratio statistic and the $R^2$ index in the linear
	regression setting \cite{veall1994evaluating}.} 

Finally, likewise for the non-fuzzy case, inference on $\boldsymbol{\beta}$ and $\boldsymbol{\gamma}$ can be performed using maximum-likelihood (ML) theory \cite{denoeux2011maximum,mclachlan2004finite} and, consequently, hypothesis testing on model's parameters can be performed using fuzzy version of likelihood ratio test (e.g., see \cite{berkachy2019fuzzy,najafi2010likelihood}). {In this context, as for the maximum-likelihood theory under the EM procedure, inferential results are based on the asymptotic properties of ML-based estimators. For further details, we refer the reader to \cite{denoeux2011maximum}, \cite{mclachlan2004finite}, and \cite{berkachy2019fuzzy}. }

\section{Simulation study}

The aim of this study is twofold. First, we will evaluate the performances of EM estimators for location and precision parameters for the fuzzy beta linear model. Second, we will assess whether standard methods, such as fixed/random-effects beta linear models, perform as good as the proposed method if applied on defuzzified data. Although the EM algorithm for fuzzy data has been validated elsewhere (e.g., see \cite{de2011measuring,su2015parameter}), in the present study we have preferred to evaluate the performances of the fuzzy-EM procedure to further provide converging results. The whole simulation procedure has been performed on a (remote) HPC machine based on 16 cpu Intel Xeon CPU E5-2630L v3 1.80 GHz,16x4 GB Ram whereas computations and analyses have been done in the \texttt{R} framework for statistical analyses. \\

\textit{Design}. The design involved three factors, namely (i) $n\in\{50,100,250,500\}$, (ii) $J\in\{2,4\}$, and (iii) $H\in\{1,3\}$, which were varied in a complete factorial design, producing $K = 4\times 2\times 2 = 16$ possible combinations. For each combination, $B=5000$ samples were generated yielding to $5000\times 16 = 80000$ new data as well as an equivalent number of parameters. The true parameters of the model were fixed as follows:
\begin{align*}
	& \boldsymbol{\beta}^0 = \left\{ (-0.5,-0.81), (-0.5,-0.81,0.7,1.15) \right\}\\
	& \boldsymbol{\gamma}^0 = \left\{ (4.8),(4.8,-1.5,1.03) \right\}
\end{align*}

\textit{Procedure}. Let $n_l$, $j_k$, $h_m$ be distinct levels of factors $n$, $J$, $H$. Then, fuzzy data were generating according to the following procedure which mimics the hierarchical process underlying rating under decision uncertainty:
\begin{enumerate}
	\item[(a)] $\mathbf X_{n_l\times j_k} = [\mathbf 1_{n_l}, \mathbf X_{n_l\times j_{k-1}}]$ and $\mathbf Z_{n_l\times h_m} = [\mathbf 1_{n_l}, \mathbf Z_{n_l\times h_{m-1}}]$, where $\mathbf X_{n_l\times j_{k-1}}$ and $\mathbf Z_{n_l\times h_{m-1}}$ were drawn from $\mathcall Unif(1,5)$\\
	\item[(b)] location and precision terms were computed as follows: 
	\begin{align*}
		& \boldsymbol{\mu}_{n_l\times 1} = \text{logit}^{-1}\left\{1+\exp(\mathbf{X}_{n_l \times j_k} \boldsymbol{\beta}^0_{j_k\times 1}) \right\}\\
		& \boldsymbol{\phi}_{n_l\times 1} = \exp(\mathbf{Z}_{n_l \times h_m} \boldsymbol{\gamma}^0_{h_m\times 1})
	\end{align*}
	\item[(c)] crisp data underlying fuzzy observations were generated according to the variable dispersion beta linear model $y_i \sim \mathcall Beta(\mu_i\phi_i,\phi_i-\phi_i\mu_i)$, $i=1,\ldots,n_l$\\
	\item[(d)] fuzzy data were generated by making $\mathbf y_{n_l}$ imprecise via a two-step data-generation process \cite{quost2016clustering,su2014likelihood}. First, spread components were generated as $\mathbf s_{n_l}\sim\mathcall Gamma(1.025,0.001)$. Second, modes were generated by $m_i \sim \mathcall Beta(y_is_i,s_i - s_iy_i)$, $i=1,\ldots,n_l$\\
	\item[(e)] parameters $\boldsymbol{\beta}_{j_k}$ and $\boldsymbol{\gamma}_{h_m}$ were estimated using four methods: 
	\begin{itemize}
		\item fEM: expectation-maximization estimators for the fuzzy case.\\
		\item dML: maximum-likelihood estimators on two type of defuzzified data computed using the \textit{centroid method} $y_i^* = \int_{0}^{1} y ~\fuzzyset{y_i}(y) ~dy$ and the \textit{first-maximum method} $y_i^* = \sup_{y\in (0,1)}\left\{\fuzzyset{y_i}(y)\right\}$ for $i=1,\ldots,n_l$. The ML procedure implemented in the \texttt{R} library \texttt{betareg} has been used in this case \cite{zeileis2010beta}.\\
		\item dREML: restricted maximum-likelihood estimators on defuzzified data obtained by treating fuzzy sets $\fuzzyset{y_1},\ldots,\fuzzyset{y_n}$ as random effects. In particular, for each observation $i=1,\ldots,n_l$, extremes of $\alpha$-sets $y^*_i = [\inf(\mathbf y^c_i),\sup(\mathbf y^c_i)]$ were used, with $\mathbf y^c_i = \{y\in (0,1): \fuzzyset{y_i}(y)\geq\alpha\}$ being the set obtained by cutting the $i$-th fuzzy data at $\alpha = [0.01,0.35,0.7]$. In this case, estimates have been performed using the \texttt{R} library \texttt{glmmTMB} for random-effects beta linear models \cite{brooks2017glmm}.
	\end{itemize}
\end{enumerate}

\textit{Outcome measures}. For each condition of the simulation design, sample results were evaluated using \textit{bias} of estimates and \textit{root mean square error}.\\

\textit{Results}. Tables \ref{tab1a}-\ref{tab1b} show the results of the simulation study with regards to averaged bias and root mean square error. For the sake of clarity, results for the cases $J=2$ and $J=4$ were reported separately. To better interpret the results, it should be noted that conditions with $H=1$ represent simplest cases where the precision term is held constant (the linear term for $\boldsymbol{\phi}$ is a simple intercept model). By contrast, conditions with $H=3$ represent those situations showing some levels of heterogeneity in the response variable (in this case the linear term for $\boldsymbol{\phi}$ contains two slopes). We first consider the conditions with $J=2$. With respect to the parameters $\boldsymbol{\hat \beta}$, all the methods showed negligible bias in estimating the location terms of the beta linear model both in the cases with $H=1$ and $H=3$. However, unlike dML and dREML, the fEM solution achieved lowest RMSE. Considering the parameters $\boldsymbol{\hat \gamma}$, dML and dREML algorithms showed worse performances when compared to fEM both in the case of low ($H=1$) and high ($H=3$) model complexity. In particular, they showed larger bias in estimating the precision terms of the beta linear model, with bias being higher with increasing model complexity ($H=3$). A similar pattern was also observed for RMSE. In particular, when compared to fEM, dML and dREML showed larger values, with severity increasing over model complexity. Interestingly, although the condition with $H=3$ represents a more complex situation with respect to parameters estimation, fEM outperformed the other methods. Results for the conditions with $J=4$ largely resemble those obtained with $H=1$. Also in this case, the fEM algorithm showed better performances over dML and dREML. Finally, for each method we also computed overall indices of over/under-estimation $r_{\boldsymbol{\beta}}$ and $r_{\boldsymbol{\gamma}}$, as the ratio between the number of positive and negative bias, and the overall percentage of over-estimation $p_{\boldsymbol{\beta}}$ and $p_{\boldsymbol{\gamma}}$. Table \ref{tab1c} reports the results along with the overall RMSE for both the arrays of parameters. In general, fEM and dML (mode) showed negligible overestimation for $\boldsymbol{\beta}^0$ whereas dML (mean) and dREML tended to overestimate the true arrays of parameters. On the contrary, when estimating $\boldsymbol{\gamma}^0$, fEM tended to overestimate the true population parameters whereas dML and dREML tended to underestimate. On the whole, fEM showed less variable and less biased estimates of $\boldsymbol{\beta}^0$ and $\boldsymbol{\gamma}^0$ than the other procedures adopted for defuzzified data.

\begin{table}[!h]
	\caption{Monte Carlo study: average bias and average root mean square errors for the arrays of parameters $\boldsymbol{\hat\beta}$ and $\boldsymbol{\hat\gamma}$ (case $J=2$). }
	\label{tab1a}
	\centering
	\begin{tabular}{lcccccccc}
		\toprule
		\multirow{2}{*}{$n$, $H$} &
		\multicolumn{2}{c}{fEM} &
		\multicolumn{2}{c}{dML (mean)} &
		\multicolumn{2}{c}{dML (mode)} &
		\multicolumn{2}{c}{dREML} \\ \cmidrule(lr){2-3} \cmidrule(lr){4-5} \cmidrule(lr){6-7} \cmidrule(lr){8-9}
		& {\textit{bias}}& {\textit{rmse}}& {\textit{bias}}& {\textit{rmse}}& {\textit{bias}}& {\textit{rmse}}& {\textit{bias}}& {\textit{rmse}}\\
		\midrule\\
		$\boldsymbol{\hat \beta}$ & & & & & & & \\
		$n=50$ & & & & & & & \\
		$\quad H=1$ & -0.001 &   0.198  &   -0.017  &     0.251  &    0.002  &     0.276 &    -0.032  &  0.459\\[0.1cm]
		$\quad H=3$ &  0.000 &   0.355  &   -0.007  &     0.387  &   -0.006  &     0.401 &    -0.063  &  0.626\\[0.2cm]
		$n=100$ & & & & & & & \\
		$\quad H=1$ & -0.001 &   0.120  &   -0.010  &     0.181  &    0.002  &     0.177 &    -0.043  &  0.412\\[0.1cm]
		$\quad H=3$ & -0.001 &   0.186  &   -0.007  &     0.220  &   -0.003  &     0.200 &    -0.051  &  0.433\\[0.2cm]
		$n=250$ & & & & & & & \\
		$\quad H=1$ &  0.000 &   0.073  &   -0.006  &     0.142  &    0.001  &     0.118 &    -0.042  &  0.341\\[0.1cm]
		$\quad H=3$ & -0.002 &   0.126  &   -0.008  &     0.173  &   -0.003  &     0.140 &    -0.054  &  0.331\\[0.2cm]
		$n=500$ & & & & & & & \\
		$\quad H=1$ & -0.001 &   0.050  &   -0.006  &     0.132  &    0.000  &     0.086 &    -0.035  &  0.289\\[0.1cm]
		$\quad H=3$ &  0.000 &   0.084  &   -0.003  &     0.134  &   -0.002  &     0.095 &    -0.042  &  0.255\\[0.4cm]
		$\boldsymbol{\hat \gamma}$ & & & & & & & \\
		$\quad H=1$ & 0.069 & 0.039 & -0.351 & 0.104 & -0.531 & 0.138 & -1.415 & 0.320\\[0.1cm]
		$\quad H=3$ & -0.025 & 0.425 & -0.177 & 0.540 & -0.241 & 0.714 & -0.059 & 1.077\\[0.2cm]
		$n=100$ & & & & & & & \\
		$\quad H=1$ & 0.048 & 0.033 & -0.408 & 0.102 & -0.608 & 0.145 & -1.204 & 0.270\\[0.1cm]
		$\quad H=3$ & -0.013 & 0.613 & -0.152 & 0.766 & -0.141 & 0.923 & -0.077 & 1.292\\[0.2cm]
		$n=250$ & & & & & & & \\
		$\quad H=1$ & 0.039 & 0.025 & -0.435 & 0.098 & -0.665 & 0.150 & -1.113 & 0.241\\[0.1cm]
		$\quad H=3$ & -0.031 & 0.655 & -0.180 & 0.835 & -0.190 & 1.177 & -0.148 & 1.723\\[0.2cm]
		$n=500$ & & & & & & & \\
		$\quad H=1$ & 0.028 & 0.018 & -0.447 & 0.097 & -0.681 & 0.149 & -1.138 & 0.241\\[0.1cm]
		$\quad H=3$ & -0.020 & 0.743 & -0.147 & 1.369 & -0.163 & 1.215 & -0.195 & 1.628\\[0.15cm]
		\hline
		\bottomrule
	\end{tabular}
\end{table}

\begin{table}[!h]
	\caption{Monte Carlo study: average bias and average root mean square errors for the arrays of parameters $\boldsymbol{\hat\beta}$ and $\boldsymbol{\hat\gamma}$ (case $J=4$). }
	\label{tab1b}
	\centering
	\begin{tabular}{lcccccccc}
		\toprule
		\multirow{2}{*}{$n$, $H$} &
		\multicolumn{2}{c}{fEM} &
		\multicolumn{2}{c}{dML (mean)} &
		\multicolumn{2}{c}{dML (mode)} &
		\multicolumn{2}{c}{dREML} \\ \cmidrule(lr){2-3} \cmidrule(lr){4-5} \cmidrule(lr){6-7} \cmidrule(lr){8-9}
		& {\textit{bias}}& {\textit{rmse}}& {\textit{bias}}& {\textit{rmse}}& {\textit{bias}}& {\textit{rmse}}& {\textit{bias}}& {\textit{rmse}}\\
		\midrule\\
		$\boldsymbol{\hat \beta}$ & & & & & & & \\
		$n=50$ & & & & & & & \\
		$\quad H=1$ & 0.001 & 0.401 &   0.000 &    0.439 &   0.001 &    0.523 & 0.000 & 0.521\\[0.1cm]
		$\quad H=3$ & 0.001 & 0.631 &   0.010 &    0.642 &   0.006 &    0.695 & 0.019 & 0.892\\[0.2cm]
		$n=100$ & & & & & & & \\
		$\quad H=1$ & 0.000 & 0.341 &   0.010 &    0.367 &  -0.001 &    0.469 & 0.038 & 0.518\\[0.1cm]
		$\quad H=3$ & 0.000 & 0.433 &   0.000 &    0.458 &   0.002 &    0.512 & 0.002 & 0.695\\[0.2cm]
		$n=250$ & & & & & & & \\
		$\quad H=1$ & 0.000 & 0.348 &  -0.001 &    0.455 &   0.000 &    0.539 & 0.001 & 0.760\\[0.1cm]
		$\quad H=3$ & 0.003 & 0.637 &   0.009 &    0.644 &   0.011 &    0.903 & 0.034 & 1.001\\[0.2cm]
		$n=500$ & & & & & & & \\
		$\quad H=1$ & 0.006 & 0.562 &  -0.003 &    0.545 &   0.001 &    0.669 & 0.000 & 0.974\\[0.1cm]
		$\quad H=3$ & 0.001 & 0.461 &   0.006 &    0.557 &   0.004 &    0.612 & 0.011 & 0.756\\[0.4cm]
		$\boldsymbol{\hat \gamma}$ & & & & & & & \\		
		$n=50$ & & & & & & & \\
		$\quad H=1$ & 0.090 &  0.040 &      -0.308  &   0.094 &      -0.480  &   0.139 &   -1.279 &     0.294\\[0.1cm]
		$\quad H=3$ & 0.044 &  0.601 &      -0.066  &   0.794 &      -0.035  &   0.908 &    0.020 &     1.264\\[0.1cm]
		$n=100$ & & & & & & & \\
		$\quad H=1$ & 0.075 &  0.034 &      -0.374  &   0.096 &      -0.582  &   0.142 &   -1.198 &     0.268\\[0.1cm]
		$\quad H=3$ & 0.051 &  0.732  &  0.038  &   1.260 &      -0.009  &   1.350 &   -0.143 &     1.593\\[0.1cm]		
		$n=250$ & & & & & & & \\
		$\quad H=1$ & 0.041 &  0.024 &      -0.421  &   0.094 &      -0.643  &   0.149 &   -1.140 &     0.246\\[0.1cm]
		$\quad H=3$ &    -0.035 &  0.753 &      -0.187  &   1.110 &      -0.200  &   1.198 &   -0.163 &     1.749\\[0.1cm]		
		$n=500$ & & & & & & & \\
		$\quad H=1$ & 0.030 &  0.018 &      -0.435  &   0.094 &      -0.684  &   0.151 &   -1.158 &     0.246\\[0.1cm]
		$\quad H=3$ &    -0.015 &  1.016 &      -0.132  &   1.073 &      -0.158  &   1.533 &   -0.211 &     2.556\\[0.15cm]
		\hline
		\bottomrule
	\end{tabular}
\end{table}

\begin{table}[!h]
	\caption{Monte Carlo study: overall ratios $r$ between over and under-estimation, percentage $p$ of over-estimation, and overall root mean square error. Note that all the indices were computed over $B=5000$ replications.}
	\label{tab1c}
	\centering
	\resizebox{12cm}{!}{
		\begin{tabular}{lccccccccccccc}
			\toprule
			\multirow{2}{*}{} &
			\multicolumn{3}{c}{fEM} &
			\multicolumn{3}{c}{dML (mean)} &
			\multicolumn{3}{c}{dML (mode)} &
			\multicolumn{3}{c}{dREML} \\ \cmidrule(lrr){2-4} \cmidrule(lrr){5-7} \cmidrule(lrr){8-10} \cmidrule(lrr){11-13}
			& {$r$}& $p$ & $\text{rmse}_{\text{ov}}$ & {$r$}& $p$ & $\text{rmse}_{\text{ov}}$ & {$r$}& $p$ & $\text{rmse}_{\text{ov}}$ & {$r$}& $p$ & $\text{rmse}_{\text{ov}}$ \\
			\midrule\\
			$\boldsymbol{\hat \beta}$ & 1.002 & 50.1 & 0.313  & 1.244 &    55.3 & 0.358 &   1.008 &    50.2 & 0.401 & 1.161 & 53.4 & 0.579\\[0.1cm]
			$\boldsymbol{\hat \gamma}$ & 1.388 & 44.0 & 0.361 &   0.362 &    77.8 & 0.533 &  0.351 &    79.1 & 0.636 & 0.390 & 78.4 & 0.938\\[0.2cm]
			\hline
			\bottomrule
		\end{tabular}
	}
\end{table}

\section{Applications}

In this section we will illustrate the application of the fuzzy beta model to two empirical studies involving fuzzy ratings data collected using two types of fuzzy rating scales. In particular, the first application concerns the analysis of risk-taking behavioral data collected by means of an indirect fuzzy rating approach \cite{calcagni2014dynamic}. By contrast, the second application is about the analysis of customer satisfaction data collected by means of a direct fuzzy rating approach \cite{de2014fuzzy}. {Note that the applications are chosen to review two of the most common methods for fuzzy scaling, namely fuzzy indirect (case study 1) and direct rating scales (case study 2).}

\subsection{Case study 1: Reckless-driving and risk-taking behavior in young drivers}

Reckless-driving among young people is one of the major cause of mortality and injuries worldwide \cite{toroyan2013global}. It is a very complex phenomenon involving a number of human and non-human factors, such as personality, cognitive styles, social context (e.g., family, peer group), infrastructures (e.g., roads, light), and cultures  (e.g., see \cite{biccaksiz2016impulsivity,mcnally2014re,scott2015psychosocial,taubman2010attitudes}). Several studies have recognized the role of subjective factors like sensation-seeking, normlessness, anxiety, aggressiveness, driving attitudes in determining risky behaviors \cite{biccaksiz2016impulsivity}. Researchers have also assessed the contribution of parenting styles and peer relations to young drivers' intention to take risks \cite{taubman2010attitudes, taubman2014meaning}. Because of its characteristics, assessing reckless-driving behaviors is a typical situation where self-reported measures can show some levels of decision uncertainty, which cannot appropriately be analysed using final crisp responses only. In this application, we consider a set of models where reckless-driving behaviors (\texttt{rdb}) were linearly predicted by the use of substances (\texttt{drugs}), driving anger (\texttt{anger}), and family climate (\texttt{fcrs}). In particular, we hypothesized that both the use of substances and driving anger would linearly increase the self-reported reckless-driving behaviors whereas family climate would instead acts by decreasing the amount of risky behaviors. Moreover, we also assessed whether the dispersion of fuzzy ratings data varied as a function of participants' characteristics such as gender (\texttt{sex}) and frequency of driving (\texttt{driving\_frequency}). \\

\begin{figure}[h!]
	\centering
	\resizebox{10cm}{!}{
		\input{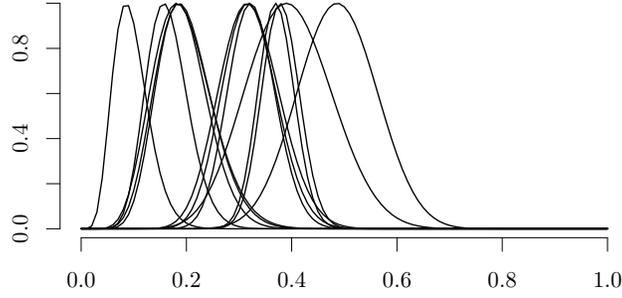}}
	\caption{Case study 1: Beta fuzzy responses on the \texttt{RDB} variable for a subsample of participants.}
	\label{fig1}
\end{figure}

\textit{Data and measures}. A questionnaire survey was carried out on $n=69$ young drivers in Trentino region (north-est of Italy). Of these, 31\% were women with mean age of 18.27 years (SD=0.56). All participants were young drivers with an average of driving experience of 12 months since receipt of their driver's license. About 73\% of them drove frequently during the week, 26\% drove once a week. The survey consisted of 24 items from three self-reported questionnaires: (i) the Reckless Driving Behavior Scale (RDB) \cite{mcnally2014re} used to assess those behaviors that increase the probability of a vehicle crash due to driving under the influence of substances ({drugs}), extreme motorsport behaviors ({extreme}), and speeding/steering behaviors ({positioning}); (ii) a short version of the Driving Anger Scale (DAS) \cite{deffenbacher1994development}, adopted to assess driving angers provoked by someone else's behaviors like slow driving and discourtesy; (iii) a simplified version of the Family Climate Road Safety (FCRS) questionnaire \cite{taubman2013family}, adopted to evaluate the role of parents in teens' safe driving, especially with regards to communication, monitoring, and parents' messages. Questionnaires were administered using DYFRAT \cite{calcagni2014dynamic}, a computerized fuzzy rating scale which adopts the mouse-tracking methodology \cite{freeman2010mousetracker} as a tool to implicitly quantify rating processes. {For each item of the survey, participants were asked to respond using a pseudo-circular rating scale with six-points (RDB), five-points (DAS), and four-points (FCRS) anchors, respectively. Participants gave their responses by mouse-clicking the chosen level of the scale. Meanwhile, we recorded the streaming $x-y$ coordinates of the computer-mouse needed to reach the chosen anchor. According to the DYFRAT methodology, the linearized histograms of the collected radians were used to compute beta fuzzy sets for each item and for each participant (fuzzy sets were obtained by means of an heuristic optimization procedure which converts histograms into fuzzy sets). Finally, modes $\mathbf m_1,\ldots,\mathbf m_n$ and precisions $\mathbf s_1,\ldots,\mathbf s_n$ were used to represent final responses and uncertainties involved during the rating process. Figure \ref{fig1} shows an example of observed beta fuzzy numbers for the \texttt{RDB} response variable.} \\

\begin{figure}[h!]
	\hspace{-1.4cm}
	\resizebox{15cm}{!}{
		\input{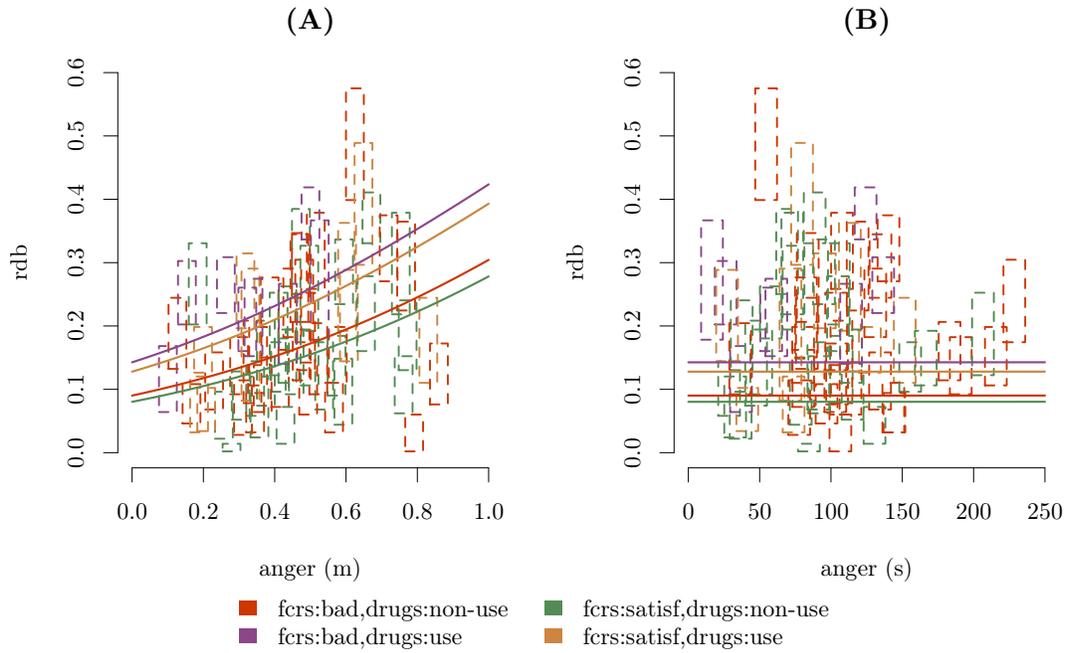}}
	\caption{Case study 1: Observed fuzzy data for \texttt{RDB} as a function of \texttt{fcrs} and \texttt{drugs} categorical predictors (colors in both panels) and the two predictors for fuzzy \texttt{anger} (panels A-B). Fitted curves correspond to model 2 in Table \ref{tab2a}. Note that rectangles represent $\alpha$-cuts of the observed fuzzy data with $\alpha=0.5$, i.e. $\mathbf y^\alpha_i = \left[\min\left(\{y\in [0,1]: \fuzzyset{y_i}(y)> 0.5\}\right),\max\left( \{y\in [0,1]: \fuzzyset{y_i}(y)> 0.5\}\right)\right]$. }
	\label{fig2}
\end{figure}

\textit{Data analysis and results}. The fuzzy response variable \texttt{rdb} was computed by aggregating the fuzzy variables extreme and positioning of the RDB questionnaire in terms of mean \cite{hanss2005applied}. Similarly, \texttt{fcrs} was obtained by aggregating the crisp responses for the variables monitoring, messages, and communications from the FCRS questionnaire. To simplify interpretation of the results, variables \texttt{drugs} and \texttt{fcrs} were made categorical using median split on their crisp responses. This yielded to two new dichotomous variables, namely \texttt{drugs} (non-use/use) and \texttt{fcrs} (bad/satisfactory family climate). By contrast, the variable \texttt{anger} entered the model as a fuzzy variable in terms of mode and precision components. First, we run a model (\textit{model 1}) where dispersion  $\boldsymbol{\phi}$ was held fixed for all participants whereas the mean $\boldsymbol{\mu}$ was modeled using the fuzzy variable \texttt{anger} and the categorical variables \texttt{fcrs} and \texttt{drugs}. Table \ref{tab2a} shows the final estimates along with their standard errors computed using the empirical observed information matrix (see Eqs. \ref{eq13}-\ref{eq14}) and Pearson's residuals. The results suggest that \texttt{rdb} increased as a function of mode components of \texttt{anger} ($\hat\beta_1=1.351$, $\sigma_{\hat\beta_1}=0.356$) whereas its precision/spread components did not affect the response variable ($\hat\beta_2=0.001$, $\sigma_{\hat\beta_2}=0.002$). Moreover, participants using drugs showed higher levels of \texttt{rdb} ($\hat\beta_4=0.458$, $\sigma_{\hat\beta_4}=0.187$) when compared to those who did not use substances. As expected, participants with satisfactory family climate showed lower levels of \texttt{rdb} ($\hat\beta_3=-0.192$, $\sigma_{\hat\beta_3}=0.440$) when compared to participants with a bad family climate. To account for heterogeneity in the response variable \texttt{rdb}, two further models were estimated, one which included the dichotomous variable \texttt{sex}, and the other which also included \texttt{driving\_frequency}. In order to evaluate models improvements, both the models were compared in terms of fuzzy likelihood-ratio test \cite{berkachy2019fuzzy}. Table \ref{tab2a} reports the results for model 2 and model 3. With regards to model 2, the likelihood-ratio test computed against model 1 reveled that \texttt{sex} improved the fit of the model ($\chi^2_{7-6=1} = 5.001$, $p=0.025$, $\text{AIC}_{\text{\tiny{model1}}} = -163.38$, $\text{AIC}_{\text{\tiny{model2}}} = -166.38$), with men showing lower levels of heterogeneity in the response variable ($\gamma_1=-1.199$, $\sigma_{\hat\gamma_1}=0.613$) when compared to women. To further analyse the variability in reckless-driving behaviors, we asked whether this varied as a function of participants' frequency of driving. To assess this hypothesis, model 3 included \texttt{driving\_frequency} as an additional term in the precision equation. The likelihood-ratio test conducted against model 2 revealed that \texttt{driving frequency} did not improve the fit of the model ($\chi^2_{8-7=1} = 1.626$, $p=0.2021$, $\text{AIC}_{\text{\tiny{model2}}} = -166.38$, $\text{AIC}_{\text{\tiny{model3}}} = -166.01$). {Thus, model 2 was retained as the final model. Overall, it showed a moderate global fit ($R^2_{\text{\tiny pseudo}} = 0.543$). The results were in line with the literature (e.g., see \cite{biccaksiz2016impulsivity}) and suggested that self-reported reckless-driving behaviors were positively associated to driving anger ($\hat\beta_1=1.485$, $\sigma_{\hat\beta_1}=0.378$) and substance use ($\hat\beta_4=0.518$, $\sigma_{\hat\beta_4}=0.235$). By contrast, family climate acted as a protective factor with satisfactory family climate being negatively associated to risky behaviors ($\hat\beta_3=-0.126$, $\sigma_{\hat\beta_3}=0.157$). Interestingly, variability of self-reported responses varied as a function of gender, with female drivers showing more variable responses than male drivers ($\hat\gamma_1=-1.999$, $\sigma_{\hat\gamma_1}=0.613$). Figure \ref{fig2} plots the predicted curves against the observed fuzzy data as a function of both continuous (panels A-B) and categorical (colors in both panels) predictors (note that only estimated modes $\boldsymbol{\hat \mu}$ were plotted for the sake of simplicity). To further investigate the role of fuzziness to predict self-reported reckless-driving behaviors (\texttt{RDB}), we contrasted the results of the final model (model 2) against the same model adapted on mean-based defuzzified response data (i.e., data without the components $\mathbf s_1,\ldots,\mathbf s_n$ for the decision uncertainty). Table \ref{tab2a-bis} shows the final estimates computed using maximum-likelihood as implemented in standard beta regression model \cite{ferrari2004beta}. The fit of the estimated model was modest ($R^2_{\text{\tiny pseudo}} = 0.243$) whereas the coefficients $\boldsymbol{\hat \beta}$ for the mean $\boldsymbol{\hat \mu}$ component were nearly similar to those obtained for the fuzzy case. By contrast, the coefficients $\boldsymbol{\hat \gamma}$ for the precisions were lower if compared to the fuzzy case. This is in line with the results of the simulation study. However, standard errors of the estimates for defuzzified data were smaller than those obtained for the fuzzy case, especially with regards to precision coefficients $\boldsymbol{\hat \gamma}$. Overall, this is coherent with the estimation approach used for fuzzy data as, in this case, standard errors were computed using score equations \eqref{eq12a}-\eqref{eq12b} which are in turn based on filtered data $(\mathbf y^*_1,\ldots,\mathbf y^*_n)$. }

\begin{table}[!h]
	\caption{Case study 1: Variable dispersion fuzzy beta models for reckless-driving behaviors. Note that categorical variables were codified using dummy coding with the following reference levels: \texttt{fcrs} (ref.: bad), \texttt{drugs} (ref.: non-use), and \texttt{driving frequency} (ref.: always). The parameters $\boldsymbol{\mu}$ and $\boldsymbol{\phi}$ were linked to the response variable using $\text{logit}(.)$ and $\log(.)$ link functions, respectively.}
	\label{tab2a}
	\centering
	\resizebox{12cm}{!}{
		\begin{tabular}{lrrr}
			\toprule
			\multirow{3}{*}{} &
			\multicolumn{3}{c}{fEM} \\ \cmidrule(lr){3-4} 
			& &{\textit{Estimate}}& {\textit{Std. Error}}\\
			\midrule\\
			\textit{Model 1} & & & \\[0cm]
			\footnotesize{Residuals quantiles: $Q1:-6.789$, $\text{Med}:-0.162$, $Q3:6.431$   } & & & \\[0.2cm]
			$\boldsymbol{\hat \mu}$ & & & \\
			\hspace{0.25cm} $\beta_0$ \footnotesize{(Intercept)} && -2.140 & 0.282 \\[0.1cm]
			\hspace{0.25cm} \texttt{anger} \footnotesize{(m)} && 1.351 & 0.356 \\[0.1cm]
			\hspace{0.25cm} \texttt{anger} \footnotesize{(s)} && 0.001 & 0.002 \\[0.1cm]
			\hspace{0.25cm} \texttt{fcrs} \footnotesize{(bad vs. satisfactory)} && -0.192 & 0.144 \\[0.1cm]
			\hspace{0.25cm} \texttt{drugs} \footnotesize{(non-use vs. use)} && 0.458 & 0.187 \\[0.1cm]
			$\boldsymbol{\hat \phi}$ && & \\
			\hspace{0.25cm} $\gamma_0$ \footnotesize{(Intercept)} && 3.330 & 0.301 \\[0.5cm]
			\textit{Model 2} & & & \\
			\footnotesize{Residuals quantiles: $Q1:-8.525$, $\text{Med}:-0.114$, $Q3:4.860$   } & & & \\[0.2cm]
			$\boldsymbol{\hat \mu}$ & & & \\
			\hspace{0.25cm} $\beta_0$ \footnotesize{(Intercept)} && -2.312 & 0.292 \\[0.1cm]
			\hspace{0.25cm} \texttt{anger} \footnotesize{(m)} && 1.485 & 0.378 \\[0.1cm]
			\hspace{0.25cm} \texttt{anger} \footnotesize{(s)} && 0.001 & 0.002 \\[0.1cm]
			\hspace{0.25cm} \texttt{fcrs} \footnotesize{(bad vs. satisfactory)} && -0.126 & 0.157 \\[0.1cm]
			\hspace{0.25cm} \texttt{drugs} \footnotesize{(non-use vs. use)} && 0.518 & 0.235 \\[0.1cm]
			$\boldsymbol{\hat \phi}$ && & \\
			\hspace{0.25cm} $\gamma_0$ \footnotesize{(Intercept)} && 4.069 & 0.472 \\[0.1cm]
			\hspace{0.25cm} \texttt{sex} \footnotesize{(female vs. male)} && -1.199 & 0.613 \\[0.5cm]
			\textit{Model 3} & & & \\[0cm]
			\footnotesize{Residuals quantiles: $Q1:-7.154$, $\text{Med}:-0.067$, $Q3:4.253$   } & & & \\[0.2cm]
			$\boldsymbol{\hat \mu}$ & & & \\
			\hspace{0.25cm} $\beta_0$ \footnotesize{(Intercept)} && -2.362 & 0.289 \\[0.1cm]
			\hspace{0.25cm} \texttt{anger} \footnotesize{(m)} && 1.551 & 0.397 \\[0.1cm]
			\hspace{0.25cm} \texttt{anger} \footnotesize{(s)} && 0.001 & 0.002 \\[0.1cm]
			\hspace{0.25cm} \texttt{fcrs} \footnotesize{(bad vs. satisfactory)} && -0.133 & 0.159 \\[0.1cm]
			\hspace{0.25cm} \texttt{drugs} \footnotesize{(non-use vs. use)} && 0.567 & 0.252 \\[0.1cm]
			$\boldsymbol{\hat \phi}$ && & \\
			\hspace{0.25cm} $\gamma_0$ \footnotesize{(Intercept)} && 3.972 & 0.485 \\[0.1cm]
			\hspace{0.25cm} \texttt{sex} \footnotesize{(female vs. male)} && -1.302 & 0.644 \\[0.1cm]
			\hspace{0.25cm} \texttt{driving\_frequency} \footnotesize{(always vs. weekend)} && 0.712 & 0.793 \\[0.1cm]
			\hline
			\bottomrule
		\end{tabular}
	}
\end{table}

\begin{table}[!h]
	\caption{Case study 1: Variable dispersion crisp beta model for reckless-driving behaviors adapted on defuzzified data. Note that categorical variables were codified as for the fuzzy beta linear models (see Table \ref{tab2a}). The parameters $\boldsymbol{\mu}$ and $\boldsymbol{\phi}$ were linked to the response variable using $\text{logit}(.)$ and $\log(.)$ link functions, respectively.}
	\label{tab2a-bis}
	\centering
	\resizebox{12cm}{!}{
		\begin{tabular}{lrrr}
			\toprule
			\multirow{3}{*}{} &
			\multicolumn{3}{c}{dML} \\ \cmidrule(lr){3-4} 
			& &{\textit{Estimate}}& {\textit{Std. Error}}\\
			\midrule\\
			\textit{Model 2} & & & \\
			\footnotesize{Residuals quantiles: $Q1:-0.3973$, $\text{Med}:0.172$, $Q3:0.636$   } & & & \\[0.2cm]
			$\boldsymbol{\hat \mu}$ & & & \\
			\hspace{0.25cm} $\beta_0$ \footnotesize{(Intercept)} && -2.199 & 0.217 \\[0.1cm]
			\hspace{0.25cm} \texttt{anger} \footnotesize{(m)} && 1.284 & 0.347 \\[0.1cm]
			\hspace{0.25cm} \texttt{anger} \footnotesize{(s)} && 0.000 & 0.001 \\[0.1cm]
			\hspace{0.25cm} \texttt{fcrs} \footnotesize{(bad vs. satisfactory)} && -0.105 & 0.120 \\[0.1cm]
			\hspace{0.25cm} \texttt{drugs} \footnotesize{(non-use vs. use)} && 0.530 & 0.145 \\[0.1cm]
			$\boldsymbol{\hat \phi}$ && & \\
			\hspace{0.25cm} $\gamma_0$ \footnotesize{(Intercept)} && 3.746 & 0.251 \\[0.1cm]
			\hspace{0.25cm} \texttt{sex} \footnotesize{(female vs. male)} && -1.082 & 0.330 \\[0.5cm]
			\bottomrule
		\end{tabular}
	}
\end{table}

\subsection{Case study 2: Service quality in restaurant industry}

Service quality is an important factor to assess the performances of a company and it is of a great interest for restaurants services as well. Several research have been conducted to understand the overall effect of perceived service quality on customer satisfaction which, in turn, leads to positive consumption behaviors like revisiting and recommending the restaurant \cite{almohaimmeed2017restaurant,namkung2008highly}. A number of variables influencing dining experience have been suggested, such as food quality, menu variety, food presentation, quality of staff service, internal/external environments \cite{ha2010effects}. Last but not least, variables like prices and restaurant type (e.g., fast-food, fine restaurants) seem to be also involved in the heterogeneity of restaurant quality \cite{almohaimmeed2017restaurant}. In this application, we will consider a simple model for restaurant service quality where perceived quality of food (\texttt{food}) and staff's perception of being courteous (\texttt{employees}) were used to predict perceived service quality (\texttt{service\_quality}). In addition, we also evaluated the extent to which heterogeneity in the response variable can be accounted by price levels (\texttt{prices}) and restaurant type (\texttt{type}). \\

\begin{figure}[h!]
	\centering
	\resizebox{10cm}{!}{
		\input{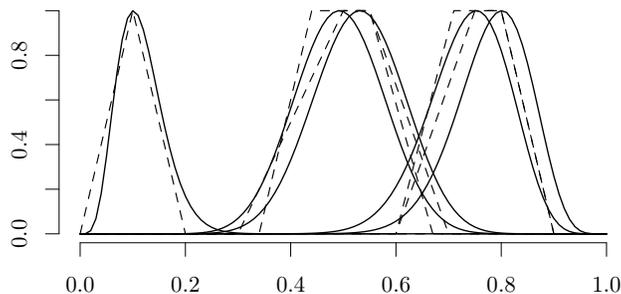}}
	\caption{Case study 2: Beta fuzzy responses (straight black lines) and trapezoidal fuzzy responses (dashed black lines) on the \texttt{service\_quality} variable for a subsample of participants.}
	\label{fig3}
\end{figure}

\textit{Data and measures}. Data were originally collected by \cite{de2014fuzzy} and refers to a survey of 14 items administered to a sample of $n=70$ customers of different age, background, and occupation. The questionnaire included two of the most important factors of restaurant quality, namely food/beverage and service quality. We considered only complete cases, i.e. cases with no missing values. This yielded to a subset of $n=49$ customers (31\% women) with modal age between 25-34 years. Informal restaurants were about 67\% of the total, 16\% of them were fine, 10\% fast-food, and 7\% were self-service restaurants. About 69\% of restaurants showed prices levels on average (about 15 Euro) whereas 31\% of them reported higher prices. {Ratings measures were collected by the authors using a Likert-type computerized fuzzy rating scale \cite{de2014fuzzy}. In this case, participants provided their responses by means of a two-step direct rating procedure. First, they were asked to draw on the graphical pseudo-continuous scale the core of the set containing the most plausible rating values. Then, conditioned on the previous choice, they were asked to draw the support of the set, which instead represents the most compatible set for the rating values. Finally, both the intervals were interpolated to form a trapezoidal fuzzy response.} For the purposes of this application, trapezoidal fuzzy numbers were converted into beta fuzzy numbers adopting a procedure minimizing the information content of the original fuzzy sets. {In general, several transformation procedures are available to this purpose \cite{nasibov2008nearest}. Here, for the sake of simplicity, we resorted to adopt the simplest approximation based on the minimization of the area under the curve between two fuzzy sets. In particular, the parameters $\{m,s\} \in (0,1)\times \mathbb{R}^+$ of the beta fuzzy set $\fuzzyset{B}(y;m,s)$ that approximates a trapezoidal fuzzy set $\fuzzyset{A}(y;a,b,c,d)$ with real parameters $a<b<c<d$ were found by minimizing the function $\delta(m,s) = |\int_{A_0} \fuzzyset{A}(y;a,b,c,d)dy - \int_{B_0} \fuzzyset{B}(y;m,s)dy|$ w.r.t. $m$ and $s$. Similarly to the first application, modes $\mathbf m_1,\ldots,\mathbf m_n$ of the beta fuzzy numbers represent final participants' responses whereas their precisions $\mathbf s_1,\ldots,\mathbf s_n$ model the decision uncertainty occurred during the rating process. Figure \ref{fig3} shows an example of transformed beta fuzzy numbers along with the associated observed trapezoidal fuzzy sets for the \texttt{service\_quality} response variable.} \\

\begin{figure}[h!]
	\hspace{-1.4cm}
	\resizebox{15cm}{!}{
		\input{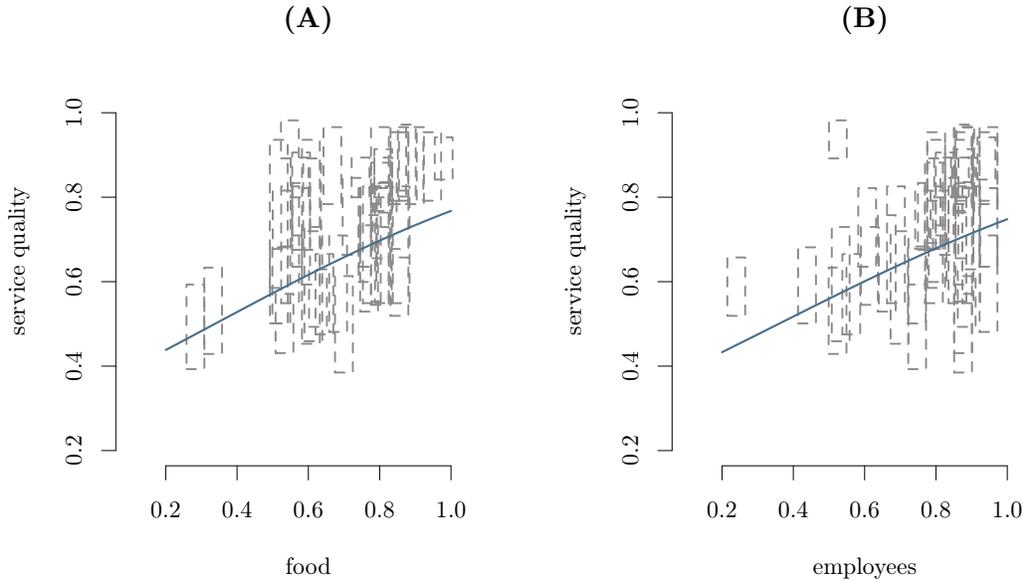}}
	\caption{Case study 2: Observed fuzzy data for \texttt{service\_quality} as a function of the two continuous predictors \texttt{food} (panel A) and \texttt{employees} (panel B). Fitted curves correspond to model 1 in Table \ref{tab2b}. Note that rectangles represent $\alpha$-cuts of the observed fuzzy data with $\alpha=0.5$, i.e. $\mathbf y^\alpha_i = \left[\min\left(\{y\in [0,1]: \fuzzyset{y_i}(y)> 0.5\}\right),\max\left( \{y\in [0,1]: \fuzzyset{y_i}(y)> 0.5\}\right)\right]$. }
	\label{fig4}
\end{figure}

\textit{Data analysis and results}. To take into account decision uncertainty for predictors, in this case fuzzy variables \texttt{food} and \texttt{employees} were defuzzified using the centroid method before entering the model. A first model was run including \texttt{food} and \texttt{employees} as predictors for the mean and \texttt{type} and \texttt{price} for precision components. Table \ref{tab2b} shows the final estimates along with their standard errors. {Overall, the model showed a moderate global fit ($R^2_{\text{\tiny pseudo}} = 0.478$).} As expected, \texttt{service\_quality} increased as a function of \texttt{food} ($\hat\beta_1=1.870$, $\sigma_{\hat\beta_1}=0.856$) and \texttt{employees} ($\hat\beta_2=1.501$, $\sigma_{\hat\beta_2}=0.718$). This is in line with previous studies on restaurant quality, suggesting that a higher perceived quality of food and staff predicts the perceived quality of restaurant services. In addition, the dispersion component of the model varied as a function of \texttt{type}, with informal restaurants showing higher heterogeneity in \texttt{service\_quality} ($\gamma_1=1.084$, $\sigma_{\hat\gamma_1}=1.055$) then self-service restaurants. The same applied for fine restaurants ($\gamma_3=0.865$, $\sigma_{\hat\gamma_3}=1.158$) whereas fast-food showed decreasing levels of variability in \texttt{service\_quality} ($\gamma_2=-0.232$, $\sigma_{\hat\gamma_2}=1.124$) when compared to self-service restaurants. Interestingly, variability in \texttt{service\_quality} is positively associated to \texttt{price} ($\gamma_4=1.802$, $\sigma_{\hat\gamma_4}=1.239$), with high-priced restaurants being more homogeneous in terms of perceived quality. This indicates that \texttt{service\_quality} was not homogeneous over normal-priced restaurants and further covariates like internal/external atmospherics or image of restaurant might instead be needed to further account for such differences \cite{almohaimmeed2017restaurant,ha2010effects}. {Figure \ref{fig4} plots the predicted curves against the observed fuzzy data as a function of both continuous predictors (note that only estimated modes $\boldsymbol{\hat \mu}$ were plotted for the sake of simplicity). As for the previous case study, we compared the fuzzy beta model with the beta regression model adapted on mean-based defuzzified data. The fit of the estimated model was modest ($R^2_{\text{\tiny pseudo}} = 0.189$). Table \ref{tab2b-bis} shows the final estimates computed using maximum-likelihood as implemented in standard beta regression model \cite{ferrari2004beta}. In line with the results from the simulation study, the estimated coefficients $\boldsymbol{\hat \gamma}$ for the precision component $\boldsymbol{\hat \phi}$ were lower than those for the fuzzy case and showed smaller standard errors. As stated previously, this can be interpreted in light of the estimation method used for the fuzzy case. }

\begin{table}[!h]
	\caption{Case study 2: Variable dispersion fuzzy beta models for service quality in restaurant industry. Note that categorical variables were codified using dummy coding with the following reference levels: \texttt{type} (ref.: self-service), \texttt{price} (ref.: high). The parameters $\boldsymbol{\mu}$ and $\boldsymbol{\phi}$ were linked to the response variable using $\text{logit}(.)$ and $\log(.)$ link functions, respectively.}
	\label{tab2b}
	\centering
	\resizebox{12cm}{!}{
		\begin{tabular}{lrrr}
			\toprule
			\multirow{3}{*}{} &
			\multicolumn{3}{c}{fEM} \\ \cmidrule(lr){3-4} 
			& &{\textit{Estimate}}& {\textit{Std. Error}}\\
			\midrule\\
			\textit{Model 1} & & & \\[0cm]
			\footnotesize{Residuals quantiles: $Q1:-5.061$, $\text{Med}:0.030$, $Q3:7.102$   } & & & \\[0.2cm]
			$\boldsymbol{\hat \mu}$ & & & \\
			\hspace{0.25cm} $\beta_0$ \footnotesize{(Intercept)} && -1.509 & 0.622 \\[0.1cm]
			\hspace{0.25cm} \texttt{food} && 1.804 & 0.875 \\[0.1cm]
			\hspace{0.25cm} \texttt{employees} && 1.696 & 0.844 \\[0.1cm]
			$\boldsymbol{\hat \phi}$ && & \\
			\hspace{0.25cm} $\gamma_0$ \footnotesize{(Intercept)} && 1.047 & 2.923 \\[0.1cm]
			\hspace{0.25cm} \texttt{type} \footnotesize{(self-service vs. informal)} && 2.267 & 1.779 \\[0.1cm]
			\hspace{0.25cm} \texttt{type} \footnotesize{(self-service vs. fast-food)} && -0.232 & 1.124 \\[0.1cm]
			\hspace{0.25cm} \texttt{type} \footnotesize{(self-service vs. fine)} && 0.865 & 1.158 \\[0.1cm]
			\hspace{0.25cm} \texttt{price} \footnotesize{(high vs. on average)} && 1.802 & 1.239 \\[0.5cm]
			\hline
			\bottomrule
		\end{tabular}
	}
\end{table}

\begin{table}[!h]
	\caption{Case study 2: Variable dispersion crisp beta model for reckless-driving behaviors adapted on defuzzified data. Note that categorical variables were codified as for the fuzzy beta linear models (see Table \ref{tab2b}). The parameters $\boldsymbol{\mu}$ and $\boldsymbol{\phi}$ were linked to the response variable using $\text{logit}(.)$ and $\log(.)$ link functions, respectively.}
	\label{tab2b-bis}
	\centering
	\resizebox{12cm}{!}{
		\begin{tabular}{lrrr}
			\toprule
			\multirow{3}{*}{} &
			\multicolumn{3}{c}{dML} \\ \cmidrule(lr){3-4} 
			& &{\textit{Estimate}}& {\textit{Std. Error}}\\
			\midrule\\
			\textit{Model 1} & & & \\[0cm]
			\footnotesize{Residuals quantiles: $Q1:-0.738$, $\text{Med}:-0.098$, $Q3:0.801$   } & & & \\[0.2cm]
			$\boldsymbol{\hat \mu}$ & & & \\
			\hspace{0.25cm} $\beta_0$ \footnotesize{(Intercept)} && -1.316 & 0.358 \\[0.1cm]
			\hspace{0.25cm} \texttt{food} && 1.597 & 0.405 \\[0.1cm]
			\hspace{0.25cm} \texttt{employees} && 1.484 & 0.383 \\[0.1cm]
			$\boldsymbol{\hat \phi}$ && & \\
			\hspace{0.25cm} $\gamma_0$ \footnotesize{(Intercept)} && 1.924 & 0.894 \\[0.1cm]
			\hspace{0.25cm} \texttt{type} \footnotesize{(self-service vs. informal)} && 0.924 & 0.843 \\[0.1cm]
			\hspace{0.25cm} \texttt{type} \footnotesize{(self-service vs. fast-food)} && -0.394 & 0.987 \\[0.1cm]
			\hspace{0.25cm} \texttt{type} \footnotesize{(self-service vs. fine)} && 0.01 & 0.953 \\[0.1cm]
			\hspace{0.25cm} \texttt{price} \footnotesize{(high vs. on average)} && 0.842 & 0.431 \\[0.5cm]
			\hline
			\bottomrule
		\end{tabular}
	}
\end{table}

\section{Conclusions}

In this article we developed a statistical approach to deal with bounded continuous ratings data in the case of non-random uncertainty. In particular, beta fuzzy numbers were adopted to represent ratings data subject to decision uncertainty and the beta regression framework was used to model the random counterpart of the overall rating process. The fuzzy component of the data was then used to estimate the latent and non-fuzzy characteristics of the raters population. Parameters estimation was performed by maximum likelihood using a version of the Expectation-Maximization algorithm generalized for the case of fuzzy data. A simulation study and two real applications were used to highlight the characteristics of the proposed approach. The simulation study revealed that the fuzzy beta linear model showed more accurate results over a set of standard methods which can be applied in the case of fuzzy data. The applications showed how the proposed method can be adopted in real cases involving ratings data represented as fuzzy numbers. 

A nice advantage of the proposed approach is its simplicity and flexibility in dealing with fuzzy data. Indeed, as it encapsulates decision rating uncertainty directly in $\mathbf{\tilde y}$, the fuzzy beta linear model does not require the extension of its parametric structure to account for modes $\mathbf m$ and precisions $\mathbf s$ of beta fuzzy data. Indeed, in the current proposal, the model's parameters $\boldsymbol{\theta} = \{\boldsymbol{\beta},\boldsymbol{\gamma}\}$ are not represented as fuzzy numbers. Consequently, parameters estimation and inference can still be performed using the asymptotic properties of maximum likelihood theory. In this setting, the fuzzy beta model recovers the parameters of the underlying rating process $Y\sim f_Y(y;\boldsymbol{\theta})$ by filtering the imprecise data in terms of expectation $\Expp{\theta}{Y|{\tilde y}}$. However, although this constitutes an important advance, it should be noticed that fuzziness of the data is not propagated to the output of the model. Indeed, as a consequence of the averaging approach upon which the fuzzy-EM is based, the predictions $\mathbf{\hat y}$ of the fuzzy beta model are non-fuzzy and they are made at the level of the underlying crisp rating mechanism $f_Y(y;\boldsymbol{\theta})$. This may limit the use of this approach in some circumstances, for example when researchers are interested in forecasting fuzziness based on current fuzzy data, or rather when components of fuzzy data play a different role in predicting the variable response (e.g., modes and precisions/spreads of the outcome variable interact with those of the explanatory variables in some way). In all these cases, different statistical approaches may instead be preferred, {like those based on full-likelihood approaches for imprecise data (e.g., see \cite{denoeux2014likelihood,kanjanatarakul2016prediction}) or min-max based estimation methods (e.g., see \cite{guillaume2020min}).}

Various possible extensions of our approach can be considered in future works. For instance, fuzzy beta model involving fuzzy data with different shapes simultaneously (e.g., beta, triangular, trapezoidal) would offer a way to deal with more complex scenarios. Another future generalization which might be interesting to investigate is the case where fuzziness in ratings data is coupled with random uncertainty which varies as a function of subgroups in the model (like for CUBE models, see \cite{piccolo2019class}). This would offer the opportunity to further decompose the overall uncertainty in ratings responses in terms of different and possibly interacting components underlying participants' rating processes. Finally, an attractive extension of the current approach would consider the case of multivariate fuzzy beta models where joint fuzzy sets do not obey to the product rule (see Eq. \ref{eq3b}). In this case, a fuzzy copula representation may instead be used to formally represent the joint fuzziness information (e.g., see \cite{ranjbar2017copula}).

\vspace{1.5cm}
\noindent\textbf{Acknowledgments}. The first author's thanks are due to Dr. Andrea Spirito for his helpful suggestions on an earlier draft of this manuscript.

\clearpage
\bibliographystyle{plain}
\bibliography{biblio}

\end{document}